\documentstyle[12pt]{article}

\begin{document}

\begin{center}{{\bf
Domain walls, $Z(N)$ charge and $A_0$ condensate: \\
a canonical ensemble study}
\footnote{Work supported by Bundesministerium f\"ur Wissenschaft,
Forschung und Kunst of Austria.}  \\
\vglue 1.0cm

O.~Borisenko, M.~Faber  \\
\vglue 0.2cm
\baselineskip=14pt
{\it Institut f\"ur Kernphysik,  Technische Universit\"at Wien,} \\
\baselineskip=14pt
{\it Wiedner Hauptstr. 8-10, A-1040 Vienna, Austria} \\
\vglue 0.4cm

G.~Zinovjev \\
\vglue 0.2cm
\baselineskip=14pt
{\it Institute for Theoretical Physics, National Academy of Sciences
of Ukraine, Kiev 252143, Ukraine}\\
\vglue 0.4cm

K.~Petrov \\
\vglue 0.2cm
\baselineskip=14pt
{\it National University of Kiev, Kiev, 252143, Ukraine}}\\
\end{center}
\vglue 0.6cm

\begin{abstract}
The deconfinement phase transition is studied in the ensemble canonical with
respect to triality. Since this ensemble implies a projection to the zero
triality sector of the theory we introduce a quantity which is insensitive
to $Z(N_c)$ symmetry but can reveal a critical behaviour in the theory with
dynamical quarks. Further, we argue that in the canonical ensemble
description of full QCD there exist domains of different $Z(N_c)$ phases
which are degenerate and possess normal physical properties.
This contradicts the predictions of the grand canonical ensemble.
We propose a new order parameter to test the realization of the discrete
$Z(N_c)$ symmetry at finite temperature and calculate it for the case
of $Z(2)$ gauge fields coupled to fundamental fermions.
\end{abstract}

\newpage

\section{Introduction}

Ever since the end of the seventies there has been increasing
interest towards discrete global and local gauge symmetries.
In 1978 G. 't~Hooft \cite{hooft} and G.~Mack \cite{mack1} conjectured
that the discrete center $Z(N_c)$ of an underlying gauge group $G$ can be
of crucial importance for quark confinement. Two years later the $Z(N_c)$
mechanism of confinement by means of vortex condensation was presented
\cite{mack2}. The next important step in understanding the role of discrete
symmetries was the awareness of the deep connection between the spontaneous
breaking of the global $Z(N_c)$ symmetry and the deconfinement phase
transition in pure gauge models \cite{gl}.
Recently, different aspects of global and local discrete symmetries have
been an object of numerous investigations both on the lattice and in
continuum. This article contributes to previous studies from the point
of view of the canonical ensemble (CE) description of hot gauge theories
with fundamental quarks.

We propose a new order parameter to test a realization of the discrete 
$Z(N_c)$ symmetry at finite temperature and calculate it for the case of
$Z(2)$ gauge fields coupled to fundamental fermions. It is argued
that domains of different N-ality in full QCD may exist,
are degenerate and possess normal physical properties.
What we would like to stress here is that in some important aspects the
two possible descriptions produce different results and, consequently
it would be safer to use the CE with respect to triality.

We begin over viewing the standard $Z(N_c)$ picture in QCD at finite
temperature \cite{gl}. Gluon fields are strictly periodic in time, while
quarks are anti periodic with a period being the inverse temperature.
Pure gauge theory has an exact $Z(N_c)$ global symmetry.
The gauge invariant operator, the Polyakov loop (PL),
$L_{\vec{x}} = \frac{1}{N_c} Tr \prod_{t=1}^{N_t} U_0(\vec{x},t)$
transforms under $Z(N_c)$ global transformations as
\begin{equation}
L_{\vec{x}} \rightarrow Z L_{\vec{x}}, \
Z = \exp [\frac{2\pi i}{N_c}n], \ n = 0,...,N_c-1.
\label{PLtr}
\end{equation}
\noindent
The PL can be used as an order parameter to test $Z(N_c)$ symmetry in
pure gauge theory. The expectation value of the PL is interpreted as the free
energy of a probe quark $F_q$ immersed in a pure gluonic bath
\begin{equation}
<L_{\vec{x}}> = \exp (-\frac{1}{T}F_q).
\label{probeq}
\end{equation}
\noindent
Unbroken $Z(N_c)$ symmetry implies $<L>=0$ and $F_q=\infty$.
When the global $Z(N_c)$ symmetry is spontaneously broken,
$<L>$ differs from zero and $F_q$ has a finite value, i.e. it costs
only finite energy to create a single quark in the gluonic bath.
To our knowledge, the first objection against this lore was discussed
in \cite{gauss} where it was pointed out that in a system with a finite UV
cutoff the free energy should not diverge. Usually, MC and analytical
calculations are performed also with periodic boundary conditions in space
directions. It has been shown, however, that Gauss' law and this p.b.c.
in space are inconsistent unless the sum of quark and gluon colour charges
vanishes, which is impossible since they have different triality.
It was concluded that for space p.b.c. the expectation value of
the PL is not the free energy of a heavy quark.
The second trouble appears if we realize that in the spontaneously
broken phase $<L>$ may pick up $N_c$ different values corresponding
to $N_c$ equivalent minima of the free energy. Thus, $L$ can be negative
or even complex. Eq.(\ref{probeq}) tells us that the free energy could be
a complex number. This gives rise to doubts that Eq.(\ref{probeq}) 
has the proper physical interpretation. When dynamical quarks
are included the picture becomes more complicated and new troubles appear.
The fermion determinant generates loops going around the lattice a number of
times which is not a multiple of $N_c$. Such loops present a propagation of
single quarks and transform non trivially under $Z(N_c)$. This means that
dynamical quarks break $Z(N_c)$ symmetry explicitly and screen sources of
heavy quarks at any temperature. The expectation value of the PL prefers
the phase with arg$L=0$ which provides the minimum of the free energy.
Other $Z(N_c)$ phases with arg$L = \frac{2\pi i k}{N_c}, k=1,\ldots,N_c-1$ become
metastable. They possess, however, such unphysical properties as a complex
free energy or entropy \cite{kogan1}. Recently, it has been discovered that
chiral symmetry is not restored in $Z(N_c)$ phases \cite{chr}.
This led to a reexamination of degenerate $Z(N_c)$ phases and interfaces
between them in pure gauge theory \cite{hanson,kiskis}.
Although not common but the most popular opinion is that all $Z(N_c)$
phases correspond to the same physical state and the interface is unphysical.

These properties reflect the GCE description of QCD with respect to
triality. In our previous papers we introduced an ensemble canonical
with respect to triality \cite{trl} (see also \cite{versus}) and
studied the QCD phase structure within this ensemble 
\cite{preprint,epce}. This ensemble reveals the following important 
properties of QCD with dynamical quarks:

1) In the low temperature phase every state has triality zero.
In the deconfinement phase the system as a whole has zero triality.
Since all $Z(N_c)$ noninvariant variables are projected out of
the theory, the PL itself has only little meaning in the CE.
A single quark does not appear
in the spectrum. Since the Debye mass is nonzero, chromoelectric fields
of quarks are screened. To distinguish the low temperature phase from
a non-confining phase one can use the phase of the $SU(N_c)/Z(N_c)$ PL.
Since this phase is insensitive to $Z(N_c)$ transformations it
can be used also in the theory with dynamical quarks \cite{im,epce}.
The deconfinement phase transition can be seen in the CE similarly to
pure gauge theory \cite{epce}.

2) Metastable minima with unphysical properties are absent, all $Z(N_c)$
phases are degenerate just as in a pure gauge model. It follows, $Z(N_c)$
phases and corresponding domain walls may exist in full
QCD \cite{trl,preprint,epce}.

3) What is not so obvious is the behaviour of the quark condensate
in different $Z(N_c)$ phases. It was shown, for example, that chiral
symmetry is always broken in the metastable minimum of $SU(2)$ theory
and the quark condensate differs from zero in this minimum at arbitrary
high temperature \cite{chr}. The canonical partition function
demonstrates opposite behaviour: chiral symmetry is restored
in all $Z(N_c)$ phases and at the same temperature. We have shown this
in our recent paper \cite{epce} by calculating the quark condensate
in the CE in strong coupling $SU(2)$ lattice gauge theory.

It should be stressed that the canonical ensemble with respect
to the total fermion number \cite{weiss} also respects these features
since it does not violate explicitly the $Z(N_c)$ global symmetry.

One may ask at this stage how to treat pure gauge theory where dynamical
quarks are absent. The resolution is trivial if we remember that pure
gauge theory is a {\it limit} of the full theory when the quark mass goes
to infinity. This limit should be taken after the thermodynamical limit
$N \rightarrow \infty$, i.e. the pure gauge partition function is 
defined as follows 
\begin{equation}
Z^{PG} =  \lim_{m_q \rightarrow \infty} \lim_{N \rightarrow \infty}
\frac{1}{N_c} \sum_{k=1}^{N_c} \int \prod_l dU_l \prod_{x,i}
d\bar{\Psi}_x^i d\Psi_x^i  \; e^{-S_W(Z_kU_0) - S_{F}}.
\label{PGpf}
\end{equation}
\noindent
Thus, by definition a single quark does not appear in the spectrum
and nonzero triality contributions are excluded by construction. Hence,
the problem is only a careful mathematical treatment of the last equation.
The puzzle mentioned in \cite{gauss} and discussed in \cite{kiskis}
does not appear. All the information can be extracted from correlators
of PLs. In this respect the observation that the CE destroys the
clusterization property has no sense as long as the PL itself has
no meaning in the CE. If one wants to have this property and relation
between the correlator in the limit $R\rightarrow \infty$
and the expectation value of the PL, the projection to another triality
sector has to be performed. This later sector, however, does not
present the ground state of QCD. $<L>$ is a thermodynamical mixture
of a heavy-light meson, a heavy-light baryon and so on (see for details,
\cite{preprint}). The correlator therefore gives
the free energy of these states placed at distance $R$.

Two questions are proper here. The first one concerns a suitable
quantity which could characterize the critical behaviour of the system
if one wants to use the CE. The second one is whether there is no
contradiction between the
screening of quarks and zero triality of the whole system. Indeed, it would
naively seem that in the Debye screening phase there are no long-range
forces. If we have a single quark at position $\vec{r}$ how can this
quark feel what is going on in any other point far away from $\vec{r}$
and, thus, how may the system have zero triality? It turns out that this
later question is closely connected to the problem of detecting
$Z(N_c)$ charges at long distances.
All these problems are  the object of the next two sections.

\section{Order parameter for $Z(N_c)$ charge and domain walls}

As an ensemble canonical with respect to N-ality does not break
$Z(N_c)$-symmetry explicitly there appears the interesting question if
$Z(N_c)$-symmetry can be broken spontaneously in this ensemble and which
quantity could be used to detect nontrivial $Z(N_c)$ charges at long
distances. An operator which detects a nontrivial phase of some state in a
region enclosed by a surface $\Sigma$ is \cite{kraus2}
\begin{equation}
A(\Sigma ,C) = \frac{<F(\Sigma) W(C)>}{<F(\Sigma)><W(C)>},
\label{OPPK}
\end{equation}
\noindent
and the limits $\Sigma , C \rightarrow \infty$ have to be taken.
$W(C)$ is a space-time Wilson loop nontrivial on the $Z(N_c)$ subgroup.
$\Sigma$ may be thought as a set of plaquettes from one time-slice
frustrated by $Z(N_c)$ singular transformations in such a way that the
dual links form a closed two-dimensional surface for this time slice.
$\Sigma$ has to be chosen to enclose a line of the Wilson loop $C$
going in time direction. The action of this order parameter is based on
the Aharonov-Bohm effect: despite the absence of chromoelectric
fields a singular potential influences particles at arbitrary
long distances giving them a nontrivial phase during winding
around the solenoid. In our case a $Z(N_c)$ charge may be viewed
as a kind of such singular solenoid placed at some space position.
To get this configuration we should take the Wilson loop $C$ and consider
the limit $C \rightarrow \infty$. Since the Wilson loop is a source of
$Z(N_c)$ charge we have the nontrivial charge and probe it via 
the ``device'' $F(\Sigma)$. A nontrivial $Z(N_c)$ charges may be detected
only in the case when the fermionic screening is suppressed relatively
to the Debye screening
%Nontrivial $Z(N_c)$ charge may be detected
%only in the case when $Z(N_c)$ symmetry is manifest.
%Otherwise $Z(N_c)$ charge is completely screened. We thus have
\begin{equation}
\lim_{\Sigma , C \rightarrow \infty} A(\Sigma , C) =
\exp \left [\frac{2\pi i}{N_c} K(\Sigma , C) \right ],
\label{znob}
\end{equation}
\noindent
where $K(\Sigma , C)$ is the linking number of the surface $\Sigma$ and loop $C$.
If the triality charge is totally screened one gets
$\lim_{\Sigma , C \rightarrow \infty} A(\Sigma , C) = 1$.

%If $Z(N)$ symmetry
%is spontaneously broken, the triality charge is totally screened
%and one gets $\lim_{\Sigma , C \rightarrow \infty} A(\Sigma , C) = 1$.

Since the fundamental PL is a source of $Z(N)$ charge
at finite temperature we can insert the correlation function
of PLs into (\ref{OPPK}) instead of the Wilson loop
\begin{equation}
A(\Sigma , R) = \frac{<F(\Sigma) L_0 L_R>}{<F(\Sigma)><L_0 L_R>}.
\label{ZNOP}
\end{equation}
\noindent

Another interpretation for the influence of $F(\Sigma)$ is the following.
$F(\Sigma)$ tries to implement a phase change at the spatial surface
$\Sigma$. If $Z(N_c)$ symmetry is spontaneously broken $F(\Sigma)$ produces a
stable interface between the volume enclosed by $\Sigma$ and the
surrounding vacuum. The different phases can be detected by a Wilson loop
and the order parameter $A(\Sigma , C)$. If the symmetry is unbroken
$F(\Sigma)$ cannot produce a stable interface and the value of the Wilson
loop is not influenced by $F(\Sigma)$ which is far apart. In the case of
the grand canonical ensemble $Z(N_c)$ symmetry is explicitly broken, there
exists a preferred phase (zero) and therefore a phase change is annihilated
after a short distance from $\Sigma$. At high and low temperatures
$\lim_{\Sigma , C \rightarrow \infty} A(\Sigma , C) = 1$ and cannot be used
as an order parameter. $Z(N_c)$ charges are completely screened by the
nonzero $N_c$-ality contributions to the partition function which are present
in the grand canonical ensemble even at high temperature and in
thermodynamical limit. As described above we expect however a nontrivial
behaviour of this order parameter in the ensemble canonical with
respect to $N_c$-ality.

To illustrate these ideas we studied a simple theory of $Z(2)$
gauge spins coupled to massless naive fermions described in the CE
by the partition function
\begin{equation}
Z = \frac{1}{2} \sum_{k= \pm 1} \sum_{s_l=\pm 1}
\int d\bar{\Psi}_x d\Psi_x \; e^{S_W + S_F}.
\label{z2ferm}
\end{equation}
\noindent
Sum over $k$ is the sum over two $N_c$-ality sectors.
We have denoted
\begin{equation}
S_W = \lambda \sum_p S_p,
\label{z2g}
\end{equation}
\noindent
where $S_p$ is a product of $Z(2)$ gauge link variables $s_l$
around plaquette and
$$
S_F = \alpha \frac{1}{2}\sum_{x,n}s_n(x)
[\bar{\Psi}_x \Psi_{x+n} - \bar{\Psi}_{x+n} \Psi_{x}]
+ \alpha \frac{1}{2}\sum_{x}s_0(x)k
[\bar{\Psi}_x \Psi_{x+0} - \bar{\Psi}_{x+0} \Psi_{x}].
$$
We fix a static gauge where all gauge spins in time direction $s_0(x)$
are set to 1 except for one time-slice which includes the frustrated
plaquettes $\Sigma$. We consider the region $\lambda \gg 1$ and
$\alpha \ll 1$ which corresponds to the weak coupling high temperature phase.
The fermionic screening is suppressed as $\alpha^{2N_t}$. The Debye
screening comes from the pure gluonic action and produces the formula for
the correlation of PLs
\begin{equation}
\ln <L_0 L_R> \propto \exp [-M_D R],
\label{z2cor}
\end{equation}
\noindent
where $M_D$ is the Debye mass. It is straightforward to calculate
$<F(\Sigma)>$ in leading order in $\alpha$
\begin{equation}
<F(\Sigma)> = \exp [-\delta S_{\Sigma} + O(\alpha^8)],
\label{z2F}
\end{equation}
\noindent
where $\delta \approx \frac{2\alpha^4 \tanh \lambda}{1+\alpha^2}$.
The main contribution comes from configurations
of the gauge fields $s_0(x)$ flipped in a volume enclosed by $\Sigma$
relatively to $s_0(x)$ outside of this volume. Since the PL $L_0$
in the origin penetrates this volume one time $L_0$ changes its sign.
The corresponding expansions in large $\lambda$ and in small $\alpha$ are
converging. Therefore, in the limit of infinite $\Sigma$ and $R$
all the corrections go to zero and we find
\begin{equation}
A(\Sigma , R) = -I.
\label{z2ZNOP}
\end{equation}
\noindent
In the region $\alpha \gg 1$ the fermionic screening can dominate
$<L_0 L_R>$ and we find $A(\Sigma , R) = I$. It follows that
$Z(2)$ may be broken in this region. We have thus shown that
$A(\Sigma , R)$ can be used in the CE as an order parameter to distinguish
the phase of broken $Z(N_c)$ symmetry from the phase where the triality
charge can be detected at long range.
The operator $<F(\Sigma)>$ introduces a $Z(N_c)$ string which becomes the
boundary of a domain wall when the $Z(N_c)$ symmetry is spontaneously
broken. This domain wall is stable in pure gauge theory but appears
to be unstable in full QCD if we treat it in the GCE. In the CE the
domain wall is stable just as in pure gauge theory.
Thus, the question of the realization of the domain
walls in QCD is only the question whether the local $Z(N_c)$ symmetry
can be spontaneously broken in the CE in the sense described here.
Application of these results to $SU(N_c)$ theory is conceptually
straightforward. Calculations however are much more complicated
and could be advanced using Monte-Carlo simulations.

\section{$A_0$ condensate in the CE}

We consider in this section another quantity which is a good
candidate for characterizing the deconfinement phase, the so-called
$A_0$ condensate. It carries other useful features 
than the order parameter proposed above. Let us perform the following
decomposition of the $SU(N_c)$ PL $V_x$ before tracing ($L_x=TrV_x$)
\begin{equation}
V_x = Z_x \bar{V}_x,
\label{zv}
\end{equation}
\noindent
where $\bar{V}_x \in SU(N_c)/Z(N_c)$ and $Z_x \in Z(N_c)$.
For the invariant measure we have in this case
\begin{equation}
\int D \mu (V) = \frac{1}{N_c} \sum_{Z} \int D \mu (\bar{V}),
\label{imzv}
\end{equation}
\noindent
where $D \mu (\bar{V})$ is the invariant measure on $SU(N_c)/Z(N_c)$ group.
Let us recall, that the invariant measure on the $SU(N_c)/Z(N_c)$ group
coincides with the $SU(N_c)$ measure up to the restriction
\begin{equation}
-\frac{2\pi}{N} \leq arg[TrV] \leq \frac{2\pi}{N}.
\label{region}
\end{equation}
\noindent
We chose the static diagonal gauge for $V_x$ gauge field matrices.
Then, $\bar{V}_x$ has the diagonal representation with
angles $\phi_k$ obeying Eq. (\ref{region}). The corresponding continuum
potential can be written in the form $\beta g(A_0 + a_0(x))$ with a constant
part $A_0$ taking values as in (\ref{region}) and a quantum fluctuating
part $a_0(x)$. This constant value may be interpreted as the $A_0$ 
condensate, see \cite{review} for a review on this question.
Our definition is, however, slightly different just because of the
restriction (\ref{region}). We have found that this definition is more
proper both for the continuum theory and for the lattice theory.
$Z$ configurations are
responsible for the disorder in the confinement phase and $A_0$ takes
values on the edge of the integration region
(e.g., $\beta gA_0 = \pm \pi$ for $SU(2)$).
In the high temperature phase, it is possible to show
that an actual value of $A_0$ is shifted from the edge of the integration
region forming a non-trivial saddle point in the $SU(N_c)/Z(N_c)$ subgroup
\cite{im} (truly, we did not prove that this saddle point survives
a transition to the continuum)\footnote{It should be mentioned that
there is no a common opinion in the literature concerning the
$A_0$ condensate. Very different points of view have been stated and
advocated during last five years, see \cite{review,im} and references
therein. We are not going to discuss here all the issues on this subject
conjecturing that a generation of $A_0$ is possible both on the lattice
and in the continuum space-time. We claim, however, that the value
of $A_0$ should be calculated after a summation over $Z(N)$ configurations
contained into the PL \cite{im}.}.

We, thus, adjust the following definitions for the $A_0$ condensate:

1. On the lattice we define it as a saddle point configuration
for the invariant integrals over the zero component gauge field
matrix. Any such saddle point, if it exists, is invariant by the
lattice construction. A physical meaning of this
definition follows from the Hamiltonian formulation of lattice QCD:
a constant saddle point is an (imaginary) chemical potential
for the global colour charge \cite{glch}. The existence of this
constant saddle point configuration was demonstrated in \cite{im}
for the Lagrangian formulation and in \cite{ham} for the Hamiltonian
formulation.

2. In the continuum the $A_0$ condensate is defined as the position
of the minimum of the effective potential obtained by perturbative
integration over gauge potentials over the vacuum $A_i=0$,
$A_0=$constant with fixed gauge. The proof of an independence
of the condensation phenomenon of the gauge fixing as well as
the independence of the physical quantities of the gauge in the theory
with nonzero condensate was done in Refs.\cite{skalozub}. Besides
the same physical interpretation as before, the condensate plays
the role of an infrared regulator of the theory. It was shown, for
example, that there are no infrared divergences in the two-loop
polarization tensor in a theory with such a condensate \cite{kalash}.

One important advantage of our definition
of $A_0$ is that it is insensitive to the $Z(N_c)$ symmetry because
$\bar{V}_x$ is invariant under $Z(N_c)$ transformation.
Hence, it may be used both in a pure gauge theory and in a theory with
dynamical fermions in both ensembles to reveal some features of
the high-temperature phase. We think that $A_0$ could be used as a
sort of ``order parameter'' to distinguish the confining from 
the non-confining phase in the CE.  
Presumably, it is not an order parameter in the exact
meaning, rather it shows preferred values of the $SU(N_c)/Z(N_c)$
configurations of the PL in different phases. In fact, this is very close
to what the adjoint PL demonstrates in the $SO(3)$ LGT at finite
temperature \cite{so3}. Even though there is no global $Z(2)$ symmetry
in the $SO(3)$ model, the phases of the adjoint PL behave differently
in the confinement and in the deconfinement phases 
as described above. The $SO(3)$ model was considered in this respect
also in \cite{review}. We have proved for the simple example of a pure
$SO(3)$ model in the strong coupling region that the phases of the adjoint
PL can indeed be used to test the deconfinement phase. In the continuum
theory in the CE the $A_0$ condensate was calculated in \cite{ols} up to
second loop order. It shows the expected behaviour at high temperatures.
The loop expansion, however does not allow to observe a critical behaviour
of the system. This is a question we address here. We are able now
to demonstrate this expected behaviour in the strong coupling $SU(2)$
LGT with Kogut-Susskind fermions. We work in the Hamiltonian formulation.
The Hamiltonian of the lattice gluodynamics in the strong coupling
approximation includes only the chromoelectric part
\begin{equation}
H = \frac{g^{2}}{2a} \sum_{links} E^{2}(l),
\label{a01}
\end{equation}
\noindent
where $E(l) = i\partial / \partial (A_{l})$ are the chromoelectric field
operators. In this approach the chromomagnetic term can be treated
perturbatively at $g^{2} \rightarrow \infty $.
The determination of the partition function
\begin{equation}
Z = \tilde{Sp} \exp (-\beta H)
\label{a02}
\end{equation}
\noindent
is connected with the summation over local gauge-invariant
states. This is reflected by the symbol $\tilde{Sp}$ in (\ref{a02}),
$\beta$ is the inverse temperature.
The corresponding physical Hilbert space is determined by
Gauss' law. After the conventional procedure one gets the partition function
of the form (see, for instance, \cite{review})
\begin{equation}
Z = \int \prod_{x}d \mu (\phi_x) \prod_{x,n} \left [
\sum_{l=0,\frac{1}{2},...}
e^{-\gamma C_l} \Omega_{l}(\phi_{x}) \Omega_{l}(\phi_{x+n}) \right ].
\label{a03}
\end{equation}
$C_l$ is here the quadratic Casimir operator,
$\gamma = \frac{\beta g^2}{2a}$. $\Omega_l$ is the character
of $l$-th irreducible representation of the $SU(2)$ group
\begin{equation}
\Omega_{l}(\phi) = \frac{\sin (2l+1)\phi}{\sin \phi}.
\label{a04}
\end{equation}
\noindent
Notice, that the invariant measure $d \mu (\phi_x)$ appeared after
the representation of the Gauss' law delta-function on the $SU(2)$ group.
Following formulae (\ref{zv})-(\ref{region}) we represent $SU(2)$
characters as
\begin{equation}
\Omega_{l}(\phi) = s^{2l}\bar{\Omega}_{l}(\phi).
\label{a05}
\end{equation}
\noindent
Here, we may choose $-\frac{\pi}{2} \leq \phi \leq \frac{\pi}{2}$ which
follows from (\ref{region}). We would like to calculate an effective
potential for the phase $\phi$ which we set constant further. In this
way, using a minimization procedure, we find a saddle point for integrals
over the $SU(2)/Z(2)$ subgroup. Substituting (\ref{a04}) and (\ref{a05})
into (\ref{a03}) and summing up over all representations of the $SU(2)$ 
group, one arrives to the effective potential of the form
\begin{equation}
V_{eff}(\phi) = 2d \ln F_0(\phi) - (2d-1)\ln \sin^{2} \phi
+ \frac{1}{N_s} \ln Z_{IM}(\phi),
\label{a06}
\end{equation}
\noindent
where $N_s$ is a number of lattice sites and we have taken into
account the measure contribution $\ln \sin^{2} \phi$ which is invariant
under $Z(2)$. $Z_{IM}$ is the partition function of the Ising model
\begin{equation}
Z_{IM} = \sum_{s_x=\pm 1}\prod_l[1 + \tanh Y s_xs_{x+n}]
\prod_x [1 + M \cos \phi s_x],
\label{a07}
\end{equation}
\noindent
where
\begin{equation}
\tanh Y = \frac{F_1(\phi)}{F_0(\phi)}
\label{a08}
\end{equation}
\noindent
and
\begin{equation}
F_0(\phi) = \Theta_2(e^{-\gamma},0) - \Theta_2(e^{-\gamma},2\phi),
\label{a09}
\end{equation}
\noindent
\begin{equation}
F_1(\phi) = \Theta_3(e^{-\gamma},0) - \Theta_3(e^{-\gamma},2\phi).
\label{a010}
\end{equation}
\noindent
$\Theta_i$ is the Jacobi theta-function. The quark contribution
\begin{equation}
Z_q = \prod_x [1 + M \cos \phi s_x]
\label{a011}
\end{equation}
\noindent
was taken as in \cite{banks} with $M = \exp(-\beta m)$.
The first two terms from (\ref{a06}) coincide 
exactly with the effective potential for pure the $SO(3)$ model
in the same approximation \cite{review}. The existence of a nontrivial
solution for $\phi$ can easily be established in this case for
$\gamma > \gamma_c$. For $\gamma < \gamma_c$ one finds
$\phi = \pm \frac{\pi}{2}$. Thus, for this case our conclusion coincides 
with the one of Ref.\cite{so3}. The third term in (\ref{a06})
establishes a difference from the $SO(3)$ theory. It is clear, however,
that this term cannot influence drastically the qualitative picture in
the deconfinement phase.
We performed numerical calculations to find  maxima of the potential
for the CE, where for the quark contribution
we should take
\begin{equation}
Z_q^{CE} = \sum_{k=0}^1 \prod_x [1 + M \cos (\phi + k\pi) s_x],
\label{a012}
\end{equation}
\noindent
whereas for the $SO(3)$ model this contribution is
\begin{equation}
Z_q^{SO(3)} = \prod_x [1 + M \cos^2 \phi ].
\label{so3q}
\end{equation}
\noindent
Detailed calculations with a chromomagnetic part of the Hamiltonian
are at the moment in progress and will be presented elsewhere.
The result for the effective potential (\ref{a06}) including the quark 
contribution is, however, qualitatively the same both for the $SO(3)$ 
model and for the $SU(2)$ model in the CE, namely $A_0$ forms a nontrivial 
saddle point above some critical value. Below this value $\phi$ is 
$\pm \frac{\pi}{2}$. Therefore, we established that $A_0$ 
defined in such a manner, indeed can serve as an order parameter 
to probe phase transitions in a gauge theory with fundamental 
matter fields.

As a last point we would like to mention an application
of our result. We have already stressed that the $A_0$ condensate
has the meaning of an imaginary chemical potential for the global
colour charge. Thus, if the condensate appears above some critical
point, a spontaneous generation of nonzero colour charge might
happen in the quark sector of the theory. 
This charge should have the same absolute value 
in all $Z(N_c)$ phases. Since it must be zero in the confinement region
this charge could be an exact order parameter for full QCD.
If this phenomenon really happens it could lead to significant
improvement of our knowledge of the quark-gluon plasma phase.


\begin{thebibliography}{99}
\baselineskip=14pt


\bibitem{hooft} G.~'t Hooft, Nucl.Phys. B138 (1978) 1.
%
\bibitem{mack1} G.~Mack, Phys.Left. 78B (1978) 263.
%
\bibitem{mack2} G.~Mack, V.B.~Petkova, Ann. of Phys. 125 (1980) 117.
%
\bibitem{gl} L.~D.~McLerran and B.~Svetitsky, Phys.Rev. D24
(1981) 450; B.~Svetitsky and L.G.~Yaffe, Nucl.Phys. B210 (1982) 423;
J.~Kuti, J.~Polonyi, K.~Szlachanyi, Phys.Lett. 98B (1981) 199.
%
\bibitem{gauss} E.~Hilf, L.~Polley, Phys.Lett. B131 (1983) 412.
%
\bibitem{kogan1} V.~M.~Belyaev, Ian~I.~Kogan, G.~W.~Semenoff and N.~Weiss,
Phys. Lett. B 277 (1992) 331; W.~Chen, M.~I.~Dobroliubov and G.~B.~Semenoff,
Phys. Rev. D46 (1992) 1223.
%
\bibitem{chr} S.~Chandrasekharan, N.~Christ, hep-lat/9509095;
M.A.~Stephanov, hep-lat/9601001.
%
\bibitem{hanson} T.H.~Hansson, H.B.~Nielsen and I.~Zahed,
Nucl.Phys. B451 (1995) 162; Ian I.~Kogan, Phys.Rev. D49 (1994) 6799.
%
\bibitem{kiskis} A.~Smilga, Ann.Phys. 234 (1994) 1;
J.~Kiskis, Phys.Rev. D51 (1995) 3781; hep-lat/9510029.
%
\bibitem{trl} M.~Faber, O.~Borisenko, G.~Zinovjev,
Nucl.Phys. B444 (1995) 563.
%
\bibitem{versus}  M.~Oleszczuk and J.~Polonyi, Canonical vs. grand
canonical ensemble in QCD, preprint TPR 92-34, 1992;
J.~Polonyi, Heavy Ion Physics 2 (1995) 123.
%
\bibitem{preprint} M.~Faber, O.~Borisenko, S.~Mashkevich, G.~Zinovjev,
Nucl.Phys. B (Proc.Suppl.) 42 (1995) 484.
%
\bibitem{epce} O.~Borisenko, M.~Faber, and G.~Zinovjev,
hep-lat/9604020, submitted to Mod.Phys.Lett.
%
\bibitem{weiss} N.~Weiss, Phys.Rev. D35 (1987) 2495.
%
\bibitem{kraus2} J.~Preskill, L.M.~Krauss, Nucl.Phys. B341 (1990) 50;
K.~Li, Nucl.Phys. B361 (1991) 437.
%
\bibitem{im} O.~Borisenko, J.~Bohacik, hep-lat/9607001,
submitted to Phys.Rev.D.
%
\bibitem{review}  O.~Borisenko, J~Boh\'a\v cik, V.~Skalozub,
Fortschr.Phys. 43 (1995) 301.
%
\bibitem{glch} O.~Borisenko, V.~Petrov, G.~Zinovjev and
J.~Boh\'a\v cik, Mod.Phys.Lett. A6 (1992) 1429;
M.~Oleszczuk, J.~Polonyi, preprint CTP-2037, 1991.
%
\bibitem{ham} O.~Borisenko, V.~Petrov, G.~Zinovjev,
Phys.Lett. B264 (1991) 166. 
%
\bibitem{skalozub} V.~Skalozub, Phys.Rev.D 50 (1994) 1150;
Int.J.Mod.Phys. A9 (1994) 4747.
\bibitem{kalash} O.~Kalashnikov, Prog.Theor.Phys. 92 (1994) 1207.
%
\bibitem{so3} S.~Cheluvaraja, H.S.~Sharathchandra, Finite temperature
properties of $SO(3)$ lattice gauge theory and their implications
for the continuum theory, preprint hep-lat/9604007.
%
\bibitem{ols} M.~Oleszczuk, Preprint TPR-94-11, 1994.
%
\bibitem{banks} T.~Banks, A.~Ukawa, Nucl.Phys. B225 (1983) 145.

\end{thebibliography}
\end{document}